\documentstyle[iopconf1]{article}

\def\beqa{\begin{eqnarray}}
\def\eeqa{\end{eqnarray}}
\def\beq{\begin{equation}}
\def\eeq{\end{equation}}

\def\half{\frac{1}{2}}

\def\mD{\mbox{\bf D}}
\def\mE{\mbox{\bf E}}
\def\mH{\mbox{\bf H}}
\def\mB{\mbox{\bf B}}
\def\mr{\mbox{\bf r}}

\def\dmunu{_{\mu\nu}}

\def\uab{^{\alpha\beta}}

\def\pa{\partial}

\let\alp=\alpha

\def\apj{{\it Ap. J.}\ }

\def\etal{{\it et al.}}

\begin{document}
\title{Large Scale Structure and Cosmological Waveguides}

\author{S Capozziello\footnote{E-mail:
capozziello@vaxsa.csied.unisa.it} and G Iovane\footnote{E-mail:
geriov@vaxsa.csied.unisa.it}}

\affil{Dipartimento di Scienze Fisiche "E.R. Caianiello", Universit\`a
di Salerno, I-84081 Baronissi (SA), Italy}

\beginabstract
A sort of gravitational waveguide effect in
cosmology could explain some anomalous phenomena which
cannot be understood by  the current
 gravitational lensing models as the existence of
"twins" objects with similar spectra
and redshifts posed on the sky with large angular distance.
Furthermore,  the huge luminosities
of quasars could be explained using filamentary or planar
cosmological structures acting as waveguides.
We describe the gravitational waveguide theory and then we
discuss possible realizations in cosmology.
\endabstract

\section{Introduction}
Besides traditional fields of astronomy and astrophysics, gravitational
lensing can be today considered a fundamental tool to investigate the large
scale structure of the universe and to  test  cosmological models. In some
sense, we can say that a sort of ``gravitational'' astronomy is coming up.
One of the most interesting characteristics of gravitational lensing is
that it acts on all scales.
It provides a great amount of cosmological and astrophysical
applications like the determination of the Hubble parameter $H_{0}$ {\it via}
the measurement of  time delay $\Delta t$ between the observed lightcurves
of multiply imaged extragalactic sources;
the possibility of weighing the mass and describing the potential of lensing
galaxies and galaxy clusters from the observation of multiply imaged
quasars, arcs and arclets.
Furthermore, the gravitational lensing
plays a leading role in searching for dark matter at large scales
 since the frequency of
multiply imaged sources (e.g. quasars) depends on the cosmological density
parameter $\Omega_{L}$ of compact objects.
Particularly promising are the multiply
macro--imaged quasars whose lensing galaxy should have a large optical depth
for lensing effects
(at least 20 objects of this kind  have been
identified till now).
The above kinds of analysis are possible if we have a model
explaining the way of forming images such as the above--mentioned arcs,
rings or simply double images and predicting the effects of the deflector.
The gravitational
lensing may be explained through the action of a weak gravitational
field on the light rays. The action of  media with
corresponding refraction index is
completely determined by the Newtonian gravitational potential
which deflects and focuses the light rays.
In optics, however, there exist
other types of devices, like optical fibers and  waveguides which use the same
deflection phenomena. The analogy with the action of a gravitational field
onto light rays may be extended considering also these other
structures. In other words, it is
 possible to suppose the existence of a sort of
gravitational waveguides  \cite{capozziello}.
 On the other hand, structures
like cosmic strings, texture and domain walls,
 which are produced at phase transiton in inflationary
models, can evolve into today observed filaments, clusters and groups
of galaxies and behave in a variety of ways with respect to
gravitational lensing effects.
For example, a filament of galaxies can be considered  a sort of
waveguide preserving total luminosity of a source, if we have locally an
effective
gravitational potential of the form $\Phi(\bf{r})\sim r^{2}$, while
the planar structures generated by the motion of cosmic strings (the so
called "wakes") can yield cosmological structures where the total
flux of light is preserved and the brightness of objects at high redshift,
whose radiation passes through such structures, appears higher
to a far observer.
In this note,
 we costruct a  waveguide model using the
paraxial approximation and consider
the effect  of gravitational systems
which, combined in filaments or in planes, results as  waveguides.
Finally, we discuss the eventual cosmological realization of
such structures and the connection with observations.

\section{The propagation of light in a weak gravitational field}
The discussion of gravitational waveguide properties can be done starting
from
the electromagnetic field theory in a gravitational field described by
the metric tensor $g\dmunu$ \cite{ehlers}. In this context,
the behaviour of the electromagnetic field, without sources,
may be described by the Maxwell equations
\beq
\label{14}
\frac{\pa F_{\alp\beta}}{\pa x^{\gamma}}+
\frac{\pa F_{\beta\gamma}}{\pa x^{\alp}}+
\frac{\pa F_{\gamma\alpha}}{\pa x^{\beta}}=0\;;
\;\;\;\;
\frac{1}{\sqrt{-g}}
\frac{\pa}{\pa x^{\beta}}\left(\sqrt{-g} F^{\alpha\beta}\right)=0\,,
\eeq
where $F\uab$ is the electromagnetic field tensor and $g$ is the determinant
of the four--dimensional metric tensor. For a static gravitational field,
these equations can be reduced to the usual Maxwell equations describing the
electromagnetic field in media where the dielectric and magnetic tensor
permeabilities are connected with the metric tensor $g\dmunu$ by the
equation
\beq
\label{16}
\varepsilon_{ik}=\mu_{ik}=-g_{00}^{-1/2}[\mbox{det}g_{ik}]^{-1/2}g_{ik}\;;
\;\;\;\;\;i,k=1,2,3\,.
\eeq
If one has an isotropic model, the metric tensor is diagonal and
the refraction index of "media" may be introduced by mimicking  the
gravitational field
$n({\bf r})=(\varepsilon\mu)^{1/2},$
(it is worthwhile to note that such a situation can be
easily reproduced in cosmology).
For weak gravitational fields,
the metric tensor components are expressed in terms of
the Newton gravitational potential $\Phi$ as
\beq
\label{19}
g_{00}\simeq 1+2\frac{ \Phi ({\bf r})}{c^{2}}\;;\;\;\;\;\;
g_{ik}\simeq -\delta_{ik}\left(1-2\frac{\Phi({\bf r})}{c^{2}}\right)\;;
\eeq
where we are assuming the weak field limit,
$\Phi/c^{2}\ll 1$, and the slow motion approximation $|v|\ll c$.
Then, due to relations (\ref{16}) and (\ref{19}), the
refraction index $n({\bf r})$   can be expressed in terms of the
gravitational potential $\Phi({\bf r})$ produced by some matter distribution.
Such a weak field situation is realized for  cosmological structures
which give rise to the gravitational lensing effects connected to several
observable phenomena (multiple images, magnification, image distorsion,
arcs and arclets).
The same scheme can be applied also to string and planar--like distributions
 of matter giving rise to  gravitational waveguide effects.

\section{The Helmoltz Equation}
A waveguide solution can be obtained by reducing the Maxwell equations in media
to the  Helmoltz scalar equations for the fields $\mE$ and $\mH$.
Let us take into account the media contribution  by the relations
\beq
\label{m5}
\mD_{\omega}(\mr)=\varepsilon (\omega,\mr)\mE_{\omega}(\mr)\,,\;\;\;\;
\mB_{\omega}(\mr)=\mu (\omega,\mr)\mH_{\omega}(\mr)\,,
\eeq
considering the single field components from Eq.(\ref{14}).
The subscript $\omega$ means the Fourier amplitudes of the fields,
i.e.
\beq
\label{m6}
\mE(\mr,t)=\int \mE_{\omega}(\mr)e^{-i\omega t}d\omega\,,
\eeq
and analogously for $\mD, \mB, \mH$.
Then, taking the Fourier transforms with respect to the time variable,
and with a little algebra on the Maxwell equations in vectorial form,
we get
\beq
\label{m16}
\triangle \mE_{\omega}(\mr)+\frac{\omega^2}{c^2}n^{2}(\omega,\mr)
\mE_{\omega}(\mr)=-
\nabla\left[\frac{\mE_{\omega}(\mr)
\nabla\varepsilon(\omega,\mr)}{\varepsilon(\omega,\mr)}\right]\,.
\eeq
where, as above,
$ n^2(\omega,\mr)=\mu(\omega,\mr)\varepsilon(\omega,\mr),$
is the refractive index, and $\triangle$ is the ordinary Laplace operator
which can be used in the weak field approximation which we are considering.
One can neglect the term in the rhs of Eq.(\ref{m16}) if it is
much less than both terms in the lhs of the same relation. In
fact, since $\triangle=\nabla\cdot\nabla$ for distances of an order of the
light wavelength $\lambda$, both terms in the lhs of
Eq.(\ref{m16}) (independently of the light polarization) are of the order
\beq
\label{m18}
|\triangle\mE_{\omega}(\mr)|\sim\lambda^{-2}E_{\omega}(\mr)\,,
\;\;\;\;\;
\frac{\omega^{2}}{c^2}n^2(\omega,\mr)\mE_{\omega}(\mr)
\sim\lambda^{-2}E_{\omega}(\mr)\,.
\eeq
The term depending on the light polarization interaction for the same
distances is of the order
\beq
\label{m19}
\nabla\left[\frac{\mE_{\omega}(\mr)
\nabla\varepsilon(\omega,\mr)}{\varepsilon(\omega,\mr)}\right]\sim
\lambda^{-2}\frac{\delta\varepsilon}{\varepsilon}E_{\omega}(\mr)\,,
\eeq
where $\delta\epsilon$ is the change of the dielectric permeability for
distances of the order of wavelength $\lambda$.
Comparing Eqs.(\ref{m18}) and (\ref{m19}), we conclude that for
${\displaystyle \frac{\delta \varepsilon}{\varepsilon}\ll 1},$
we can neglect the term depending on the light polarization interaction with
respect to the other two terms. In this approximation, we get the scalar
Helmholtz equation for all the decoupled components of the electric vector
field, i.e.
\beq
\label{m21}
\triangle \mE_{\omega}(\mr)+\frac{\omega^2}{c^2}n^2(\omega,\mr)
\mE_{\omega}(\mr)=0\,.
\eeq
The same holds for magnetic vector field.
If one has a solution of the Helmholtz equation $\mE_{\omega}^{(0)}(\mr)$,
either exact or approximate one,
the influence of the light polarization interaction may be taken into account
using the Born method of iteration. In fact, the Green function given by
Eq.(\ref{m21}), $G(\mr,\mr',\omega)$,  satisfies the equation
\beq
\label{m22}
\left[\triangle+\frac{\omega^{2}}{c^2}n^2(\omega,\mr)\right]G(\mr,\mr',\omega)=
\delta(\mr-\mr')\,.
\eeq
Then the solution of the equation with the polarization term has the form
\beq
\label{m23}
\mE_{\omega}(\mr)=\mE_{\omega}^{(0)}(\mr)+\int G(\mr,\mr',\omega)
\nabla\left[\frac{\mE_{\omega}^{(0)}(\mr)
\nabla\varepsilon(\omega,\mr)}{\varepsilon(\omega,\mr)}\right]d\mr\,.
\eeq
Eq.(\ref{m22}) is equivalent in form to the equation for the Green function
of the Schr\"odinger equation, if the energy constant $E$ is equal to zero.
In fact, if we write down the Hamiltonian operator
${\displaystyle \hat{\cal H}=-\half\triangle+U(\mr),}$
with $\hbar=m=1$, and the equation for the Green function of the
Schr\"odinger equation $G_{s}(\mr,\mr',E)$ which is the matrix element of the
operator $(\hat{\cal H}-E)^{-1}$ in the coordinate representation
$G_{s}(\mr,\mr',E)=\langle\mr|(\hat{\cal H}-E)^{-1}|\mr'\rangle,$
which comes from the equation
\beq
\label{m27}
\left\{-\half\triangle+U(\mr)-E\right\}G_{s}(\mr,\mr',E)=\delta(\mr-\mr')\,.
\eeq
The comparison of this equation
with Eq.(\ref{m22}) shows that they are identical for $E=0$ with the
replacement
\beq
\label{m28}
U(\mr)=-2\frac{\omega^2}{c^2}n^2(\omega,\mr)\,,\;\;\;\;
-2G(\mr,\mr')=G_{s}(\mr,\mr',0)\,.
\eeq
Thus, we have shown that if one knows the Green function $G_{s}(\mr,\mr',E)$
of the Schr\"odinger equation for the unit mass particle moving in a
potential like that in Eq.(\ref{m28}), the Green function of the Helmholtz
equation (\ref{m22}) is given by the equality
${\displaystyle G(\mr,\mr')=-\frac{1}{2}G_{s}(\mr,\mr',E=0)\,.}$
Since the Green function for the Schr\"odinger equation are studied for many
potentials, the results obtained in quantum mechanics can be applied for our
purposes to study polarization and waveguiding effects since they are
formally identical.

\section{The gravitational waveguide model}
Let us now consider the scalar Helmholtz
equation for some arbitrary monochromatic
component of the electric field
\beq
\label{18}
\frac{\pa^{2}E}{\pa z^{2}}+\frac{\pa^{2}E}{\pa x^{2}}
+\frac{\pa^{2}E}{\pa y^{2}}+k^{2}n^{2}({\bf r})E=0\,,
\eeq
where $k$ is the wave number.
The coordinate $z$, in Eq.(\ref{18}), is  the longitudinal one,
and it can measure the space distance along the gravitational
field structure produced by a mass distribution
with an optical axis. Such a coordinate may also correspond to a distance
along the light path inside a planar gravitational field structure
produced by a planar matter--energy distribution in some regions
of the universe.
Let us consider a solution of the form
\beq
\label{1}
E=n_{0}^{-1/2}\Psi\exp\left(ik\int^{z}n_{0}(z')dz'\right)\;;
 \;\;\;\;\;n_{0}\equiv n(0,0,z)\,,
\eeq
where $\Psi(x,y,z)$ is a slowly varying spatial amplitude along the
$z$ axis,
and $\exp(iknz)$ is a rapidly oscillating phase factor.
Its clear that the beam propagation  is along the $z$ axis. We rewrite
Eq.(\ref{18}),  neglecting second order derivative in longitudinal
coordinate $z$, and obtain a Schr\"odinger--like equation for $\Psi$:
\beq
\label{2}
i\lambda\frac{\pa \Psi}{\pa \xi}=-\frac{\lambda^{2}}{2}
\left(\frac{\pa^{2}\Psi}{\pa x^{2}}+\frac{\pa^{2}\Psi}{\pa y^{2}}\right)
+\frac{1}{2}\left[n_{0}^{2}(z)-n^{2}(x,y,z)\right]\Psi\,,
\eeq
where $\lambda$ is the electromagnetic radiation wavelength and we adopt
the new variable
$\xi=\int^{z}dz'/n_{0}(z'),$
normalized with respect to the refraction index
(for our application, $n_{0}(z)\simeq 1$ so that $\xi$ coincides essentially
with $z$).
At this point, it is worthwhile to note that
if one has the distribution of the matter in the form of cylinder with
a constant (dust) density $~\rho_{0} ~$,  the gravitational potential inside
 has a  parabolic  profile providing waveguide effect
for electromagnetic radiation analogous to  optical waveguides
realized in fiber optics. In this case, Schr\"odinger--like equation
is that of two--dimensional quantum harmonic oscillator
for which the mode solutions exist in the form of Gauss--Hermite
polynomials. In the case of inhomogeneous
longitudinal dust distribution in the cylinder (that is $~\rho \,(z)\,$), the
Schr\"odinger-like equation describes the model of two-dimensional parametric
oscillator for which the mode solutions, in the form of modified Gaussian
and Gauss--Hermite polynomials, exist with parameters
determined by the density dependence on longitudinal coordinate.
As a side remark, it is interesting to stress that,
considering again Eq.(\ref{2}), the term in square brackets in the rhs
plays the role of the potential
in a usual Schr\"odinger equation; the role of Planck constant is
now assumed by $\lambda$. Since the refraction index can be expressed in terms
of the Newtonian potential when we consider the propagation of light in a
gravitational field, we can write the  potential
in (\ref{2}) as
\beq
\label{pot}
U({\bf r})=\frac{2}{c^{2}}[\Phi(x,y,z)-\Phi(0,0,z)]\,.
\eeq
The waveguide effect depends specifically on the
shape of potential (\ref{pot}):
for example,  the radiation from a remote source
does not attenuate if $U\sim r^{2}$; this situation is realized
supposing a "filamentary" or a "planar" mass distribution with constant
density $\rho$. Due to the Poisson equation, the potential inside the filament
is a quadratic function of the transverse coordinates, that is of
$r=\sqrt{x^{2}+y^{2}}$ in the case of the filament and of $r=x$ in the
case of the planar structure (obviously the light propagates in the
"remaining" coordinates: $z$ for the filament, $z,y$ for the plane).
In other words, if the radiation, travelling from some source, undergoes
a waveguide effect, it does not attenuate like $1/R^{2}$ as usual,
but it is, in some sense conserved; this fact means that the source
brightness will turn out to be much stronger than the brightness of analogous
objects located at the same distance (i.e. at the same redshift
$Z$) and the apparent energy released by the source will be anomalously large.
To fix the ideas, let us estimate how the  electric field (\ref{1})
propagates into an ideal  filament whose internal potential is
\beq
\label{internal}
U(r)=\frac{1}{2}\omega^{2}r^{2}\,,\;\;\;\;\;\;
\omega^{2}=\frac{4\pi G \rho}{c^{2}}
\eeq
where $\rho$ is constant and $G$ is the Newton constant.
 A spherical wave from a source,
$E=(1/R)\exp(ikR),$
can be represented
in the paraxial approximation as
\beq
\label{4}
E(z,r)=\frac{1}{z}\exp\left(ikz+\frac{ikr^{2}}{2z}-
\frac{r^{2}}{2z^{2}}\right)\,,
\eeq
where we are using the expansion
\beq
\label{5}
R=\left(z^{2}+r^{2}\right)^{1/2}\approx
z\left(1+\frac{r^{2}}{2z^{2}}\right)\,,\;\;\;\;r\ll z\,.
\eeq
It is realistic to  assume $n_{0}\simeq 1$ so that $\xi=z$.
Let us consider now that the starting point of the filament
of length $L$ is at
a distance $l$ from a source shifted by a distance $a$ from the filament
axis in the $x$ direction. The amplitude  $\Psi$ of the field $E$, entering
the waveguide is
\beq
\label{6}
\Psi_{in}=\frac{1}{l}\exp
\left[\frac{ikl-1}{2l^{2}}\left((x-a)^{2}+y^{2}\right)\right]\,,
\eeq
and so we have
$ R=\left(l^{2}+y^{2}+(x-a)^{2}\right)^{1/2}.$
We can calculate the amplitude of the field at the exit of the filament
by the equation
\beq
\label{8}
\Psi_{f}(x,y,l+L)=
\int
dx_{1}dy_{1}G(x,y,l+L,x_{1},y_{1},l)\Psi_{in}(x_{1},y_{1},l)\,,
\eeq
where $G$ is the Green function of Eq.(\ref{2}). For the potential
(\ref{internal}), $G$
is the propagator of the harmonic oscillator.
The integral
(\ref{8}) is Gaussian and can be exactly evaluated:
$$\Psi_{f}=\frac{\omega l}{\omega l^{2}\cos\omega L+(l+i\lambda)\sin\omega l},$$
$$\times\exp\left(-\frac{(x^{2}+y^{2})
[(\omega l k)^{2}-\omega k(i+k l)\cot\omega l]-a^{2}\omega k(i+k l)\cot\omega
L}{2(1-ikl-ik\omega l^{2}\cot\omega l)}\right)$$
\beq
\label{10}
\times\exp\left(-\frac{2xa\omega k(1+kl)}{2\sin\omega L(1-ikl
-ik\omega l^{2}\cot\omega L)}\right)\,.
\eeq
The parameter $l$ drops out of the denominator of the pre--exponential factor
if the length $L$ satisfies the condition
$\tan\omega L=-\omega l$.
Eq.(\ref{10}) is interesting in two limits.
If $\omega l\ll 1$, we have
\beq
\label{112}
\Psi_{f}=\frac{1}{i\lambda}\exp
\left\{-\frac{l+i\lambda}{2\lambda^{2}l}\left[(x+a)^{2}+y^{2}\right]\right\}\,,
\eeq
which means that the radiation emerging from a point with coordinate
$(a,0,0)$ is focused near a point with coordinates $(-a,0,l+L)$
(that is the radius has to be of the order of the wavelength).
This means that, when the beam from an extended source is focused
inside the waveguide in such a way that, at a distance $L$,
an inverted image of the source is formed, having the very
same geometrical dimensions of the source. The waveguide ``draws" the
source closer to the observer since, if the true distance of the observer
from the source is $R$, its image brightness will correspond to that of
a similar source at the closer distance
\beq
\label{eff}
R_{eff}=R-l-L\,.
\eeq
If we do not have $\omega l\ll 1$, we get (neglecting the term $i\lambda/l$
compared with unity)
\beq
\label{13}
\Psi_{f}=\frac{\sqrt{1+(\omega l)^{2}}}{i\lambda}
\exp
\left\{-\frac{1+(\omega l)^{2}}{2\lambda^{2}}
\left[y^{2}+\left(x+\frac{a}{\sqrt{1+(\omega l)^{2}}}\right)^{2}
\right]\right\}\,,
\eeq
from which, in general, the size of the image is decreased by
a factor $\sqrt{1+(\omega l)^{2}}$. The amplitude increases by the
same factor, so that the brightness is $(R/R_{eff})$ times larger.
In the opposite limit $\omega l\gg 1$, we have
$\tan\omega L\rightarrow\infty$, so that $L\simeq \pi/\omega$, that is
the shortest focal length of the waveguide is
\beq
\label{foc}
L_{foc}=\sqrt{\frac{\pi c^{2}}{4G\rho}}\,,
\eeq
which is the length of focusing of the initial beam of light trapped by
the gravitational waveguide.
All this arguments apply if the waveguide has (at least roughly) a
cylindrical geometry.
The theory of planar waveguide is similar but we have to consider only
$x$ as transverse dimension and not also $y$.
The cosmological feasibility of a waveguide depends on the geometrical
dimensions of the structures, on the connected densities
and on the limits of applicability of the above idealized scheme.
In the next section, we shall discuss these features and the possible
candidates which could give rise to observable effects.

\section{Cosmic structures as waveguides}
The gravitational waveguide effect has the same physical reason
that has the gravitational lenses effect which is
the deflection
of light by a gravitational field acting as refractive media. However,
there are essential differences producing
specific predictions for observing the waveguide effect. The gravitational
lenses are usually considered as compact objects with strong enough
gravitational potential. The light rays deflected by gravitational lenses
move outside the matter which forms the gravitational lens itself.
The gravitational waveguide as well as optical waveguide is noncompact
long structure which may contain small matter density and deflection
of the light by each element of the structure is very small. Due
to very large scale sizes of the structure (we give an extimation below),
 the electromagnetic
radiation deflection by the gravitational waveguide occurs and,
in principle, it may be observed. We will mention, for example, a possibility
of brilliancy magnification of the long distanced objects (like quasars)
with large red shift as consequence of the waveguiding structure
existence between the object and the observer. This effect exists also
for a gravitational lens located between the object and the observer,
but the long gravitational waveguide may give huge
magnification, since the radiation propagates along the waveguide with
functional dependence of the intensity on the distance which does not
decrease as
$~\sim 1/R^2~$,
characteristic for free propagation. The gravitational lens,
being a compact object, collects much less light by deflecting the rays
to the observer than the gravitational waveguide structure transporting to
the direction of observer all trapped energy (of course, one needs to
take into account losses for scattering and absorbtion). From that
point of view, it is possible that enormous amount of radiation emitted by
quasars is only seemingly existing. The object may radiate a resonable
amount of energy but the existing waveguide structure transmits the energy
in high portion to the observer. Similar ideas, related to gravitational
lensing, were discussed in \cite{barn} but
the singular lens or even few  aligned strong lenses cannot produce
effect of many orders of magnification of brilliancy.
The waveguide effect may explain the anomalous high luminosity
observed in  quasars. In fact, quasars are objects at very high
redshift which appear almost as point sources but have luminosity
that are about one hundred times than that of a giant elliptical galaxy
(quasars have luminosity which range between $10^{38}-10^{41}$ W).
For example,
PKS 2000-330 has one of the largest known redshifts $(Z=3.78)$
with a luminosity of $10^{40}$ W.
Such a redshift corresponds to a distance of $3700$ Mpc,
if it is assumed that its origin is due to
the expansion of the universe and the Hubble constant is assumed
$H=75$km s$^{-1}$ Mpc. This means that light left the quasar when
the size of the universe was one--fifth of its present age where
no ordinary galaxies (included the super giant radio--galaxies)
are observed.
The quasars, often, have both emission and absorption lines in their spectra.
The emission spectrum is thought to be produced in the quasar itself;
the absorption spectrum, in gas clouds that have either been ejected from
the quasar or just happen to lie along the same line of sight.
The brightness of quasars may vary rapidly, within a few days or less. Thus,
the emitting region can be no larger than a few light--days, i.e.
about one hundred astronomic units. This fact excludes that quasars could
be galaxies (also if most astronomers think that quasars are extremely
active galactic nuclei).
The main question is how to connect this small size with the so high
redshift and luminosity.
For example, H.C. Arp discovered small systems of quasars and galaxies where
some of the components have widely discrepant redshifts \cite{arp}.
For this reason, quasar high redshift could be produced by some unknown
process and  not being simply of cosmological origin. This claim is very
controversial. However there is a fairly widely accepted
preliminary model which, in principle,  could unify all the forms
of activities in galaxies (Seyfert, radio, Markarian galaxies and
BL Lac objects). According to this model, most galaxies contain a compact
central  nucleus with mass $10^{7}-10^{9}$ M$_{\odot}$ and diameter $< 1 $
pc. For some reason,
the nucleus may, some times, release an amount of energy exceeding
the power of all the rest of the galaxy. If there is only little gas
near the nucleus, this leads to a sort of double radio source. If the nucleus
contains much gas, the energy is directly released as radiation and one
obtains a Seyfert galaxy or, if the luminosity is even larger, a quasar.
In fact, the brightest type 1 Seyfert galaxies and faintest quasars are
not essentially different in luminosity ($\sim 10^{38}$ W) also if the
question of redshift has to be  explained (in fact quasar are, apparently,
much more distant). Finally, if there is no gas at all near such an active
nucleus, one gets BL Lac objects. These objects are similar to quasar
but show no emission lines. However the mechanism to release  such a large
amount of energy from active nuclei or quasars is still unknown.
Some people suppose that it is connected to the releasing
of gravitational energy due to the interactions of internal components
of quasars. This mechanism is extremely more efficient than the
releasing of energy during the ordinary nuclear reactions. The necessary
gravitational  energy could be produced, for example,
as consequence of the falling
of gas in a very deep potential well as that connected with a very
massive black hole. Only with this assumption, it is possible
to justify a huge luminosity, a cosmological
 redshift and a small size for the quasars.
An alternative explaination could come from waveguiding effects.
As we have discussed, if light travels within a filamentary or a
planar structure, whose Newtonian gravitational potential is quadratic
in the transverse coordinates, the radiation is not attenuated, moreover
the source brightness is stronger than the brightness of analogous
object located at the same distance (that is at the same redshift). In other
words, if the light of a quasar undergoes a waveguiding effect,
due to some structure
along the path between it and us, the apparent energy released by the
source will be anomalously large, as the object were at a distance given by
(\ref{eff}). Furthermore, if the approximation $\omega l\ll 1$
does not hold,  the dimensions of resulting image would
be decreased by a factor
$\sqrt{1+(\omega l)^{2}}$ while the brightness would be $(R/R_{eff})^{2}$,
larger, then explaining how it is possible to obtain so large emission by
such (apparently)  small  objects. In conclusion, the existence of a
waveguiding
effect may
prevents to take into consideration exotic mechanism
in order to produce huge amounts of energy
(as the existence of a massive black hole inside a galactic core) and it
may justify why it is possible to observe so distant objects of small
geometrical size.
Another effect concerning the quasars may be directly connected with
multiple images in lensing. The waveguide effect does not disappear
if the axis of "filament" or if the guide plane is bent smoothly in space.
As in the case of gravitational lenses, we can observe "twin" images if
part of the radiation comes to the observer directly from the source, and
another part is captured by the bent waveguide. The "virtual" image
can then turn out to be brighter than the "real" one (in this case we may
deal with "brothers" rather than "twins" since parameters like,
spectra, emission periods and chemical compositions are similar but
the brightnesses are different). Furthermore, such a bending in waveguide
could explain large angular separations among the images of the same
object which cannot be explained by the current lens models (pointlike
lens, thin lens and so on).
Now the issue is: are there  cosmic structures which can furnish
workable models for waveguides? Have they to be "permanent" structures
or may the waveguide effect be accidental? For example an alignment of
galaxies
of similar density and structure, due to cosmic shear and inhomogeneity,
may be available as waveguide just for a limited interval of time?
In general both points of view may be reasonable and
here we will outline both of them.
Furthermore we have to consider the problem
of the abundance of such structures: are they  common and everywhere in
the universe or are they peculiar and
located in particular regions (and eras)?

We have to do a first remark on the densities of waveguide structures
which allow observable effects.
Considering Eq.(\ref{foc}) and introducing into it the critical
density of the universe $\rho_{c}\sim 10^{-29}$ g/cm$^{3}$ (that is the
value for which the density parameter is $\Omega=1$), we obtain
$L_{foc}\sim 5\times 10^{4}$ Mpc which is an order of magnitude larger than
the observable universe and it is completely unrealistic.
On the contrary, considering a  typical galactic
density as $\rho\sim 10^{-24}$ g/cm$^{3}$, we obtain $L_{foc}\sim 100$ Mpc,
which is a typical size of large scale structure.

However, an important issue has to be
taken into account: the absorption and the scattering of light by
 the matter inside the filament or the planar structure increase with
density and, at certain crital value the waveguide effect, can be invalidated.
For the smaller frequency
of broadcast range (due to strong dependence of the absorbtion cross
section on the electromagnetic wavelength) $~\sigma \sim \sigma _T\,
(\omega /\omega _0)^4,$ where Thomson cross-section
$~\sigma _T=6\cdot 10^{-25}~\mbox {cm}^2~$
and the characteristic atomic frequency is
$~\omega _0\sim 10^{16}~\mbox {s}^{-1},$ the ratio
$~\omega /\omega _0\ll 1\,,$ and the absorbtion is small. It means that
the absorbtion length $~L_a=m_p/\rho \,\sigma \,,$ (where the mass of
proton $~m_p~$ is approximately equal to the hydrogen atom mass) is
larger than the focusing length $~L_a<L_{foc}$ for the electomagnetic waves
of broadcast range. Thus, the magnification of electromagnetic waves
may be not masked by essential energy losses due to light absorbtion and
scattering processes.
However, no restrictions exist practically  if the radio band
 and a thickness of the structure $r>10^{14}$cm are considered.

In such a case, the
relative density change
between the background and the structure density
is valid till $\delta \rho/\rho\leq 1$ . This means that we have
to stay in a linear perturbation density regime.
By such hypotheses, practically all the observed large scale structures
like filaments, walls, bubbles and clusters of galaxies can result as
candidates for waveguiding effect if the restrictions on density,
potential and waveband are respected (in optical band, such  phenomena
are possible but the density has to be chosen with some care).
Concerning the second point of view (that is the existence of temporary
waveguiding effects), it could be related to the observation of
objects possessing anomalously large (compared with their neighbours)
angular motion velocities (an analysis in this sense
could come out in mapping galaxies with respect to their redshift and proper
velocities \cite{davis}).
Such a phenomenon could mean that one observes
not the object itself, but its image transmitted through the moving
gravitational waveguide. The waveguide itself could change its form or
it could be due to temporary alignments of lens galaxies.
In this case, the image of the object could move with essentially different
angular velocity than that of the observable neighbour objects whose light
reaches the observer directly (not throught the waveguide).
The discovery of long distanced objects with anomalous velocity
(and brightness) could support the hypothesis of gravitational waveguide
effect, while the evolution of the waveguide, its destruction or change
of the axis direction (from the orientation to the Earth) could
produce the effect of the disappearence (or the appearence) of
the observed object. For this analysis, it is crucial to consider
long period astronomical observations and deep pencil beam surveys of
galaxies and quasars.
In a forthcoming papers \cite{noi}, the authors are constructing a simulation
where given random distributions of quasars and waveguides, it is possible to
reproduce the observed luminosity function of quasar distribution
 explaining its anomalies.
Finally we want to stress that our treatment does not concern only
electromagnetic radiation:
actually
a waveguide effect could be observed
also for streams of neutrinos \cite{noi},
or other
particles which gravitationally interact with large scale structures
filaments or  planes.

\vfill

\vspace{-14pt}

\begin{thebibliography}{9}
\bibitem{capozziello}
Capozziello S \etal 1997 {\it Phys. Scrip.} {\bf 56} 315
\bibitem{ehlers}
Schneider P \etal 1992 {\it Gravitational Lenses} Berlin: Springer-Verlag
\bibitem{barn}
Barnothy J M 1965 {\sl Astron. J.} {\bf 70} 666
\bibitem{arp}
 Arp H C 1987 {\it Quasars, Redshifts and Controversies}
Berkeley: Interstellar Media, $\$ 7$
\bibitem{davis}
Davis M \etal 1982 \apj {\bf 253} 423
\bibitem{noi}
Capozziello S and Iovane G {\it in preparation}

\end{thebibliography}
\end{document}